\documentclass[conference,10pt]{IEEEtran}

\newcommand{\design}{QUILL}

\usepackage{cite}
\usepackage{amsmath,amssymb,amsfonts}

\usepackage{algorithm, algorithmic}
\usepackage{listings}
\usepackage{array}
\usepackage{graphicx}
\usepackage{textcomp}
\usepackage{xcolor}
\usepackage{orcidlink}
\usepackage{svg}
\usepackage{bm}
\def\BibTeX{{\rm B\kern-.05em{\sc i\kern-.025em b}\kern-.08em
    T\kern-.1667em\lower.7ex\hbox{E}\kern-.125emX}}

\usepackage{multirow}
\usepackage{makecell}
\usepackage{booktabs}
\newcolumntype{P}[1]{>{\centering\arraybackslash}p{#1}}
\newcolumntype{M}[1]{>{\centering\arraybackslash}m{#1}}
\usepackage[table]{xcolor}

\usepackage{tikz}
\DeclareRobustCommand*\circled[1]{%
  \tikz[baseline=(char.base)]{
    \node[
      shape=circle,
      fill=black,
      text=white,
      draw,
      inner sep=0.3pt,
      font=\bfseries
    ] (char) {#1};}}

\usepackage{subcaption}
\usepackage{verbatim}

\usepackage[font=footnotesize]{caption}
\usepackage{enumitem}
\begin{document}

\title{\design: An Algorithm–Architecture Co-Design for Cache-Local Deformable Attention}
\author{\IEEEauthorblockN{Hyunwoo Oh, Hanning Chen, Sanggeon Yun, Yang Ni, Wenjun Huang, Tamoghno Das, Suyeon Jang, \\and Mohsen Imani}
\IEEEauthorblockA{
University of California, Irvine \\
Email: \{hyunwooo, m.imani\}@uci.edu}
}

\maketitle

\begin{abstract}
Deformable transformers deliver state-of-the-art detection but map poorly to hardware due to irregular memory access and low arithmetic intensity. We introduce \textbf{\design}, a schedule-aware accelerator that turns deformable attention into cache-friendly, single-pass work. At its core, Distance-based Out-of-Order Querying (DOOQ) orders queries by spatial proximity; the look-ahead drives a region prefetch into an alternate buffer—forming a schedule-aware prefetch loop that overlaps memory and compute. A fused MSDeformAttn engine executes interpolation, Softmax, aggregation, and the final projection ($W_m''$) in one pass without spilling intermediates, while small tensors are kept on-chip and surrounding dense layers run on integrated GEMMs. Implemented as RTL and evaluated end-to-end, \design{} achieves up to \textbf{7.29$\times$} higher throughput and \textbf{47.3$\times$} better energy efficiency than an RTX~4090, and exceeds prior accelerators by \textbf{3.26–9.82$\times$} in throughput and \textbf{2.01–6.07$\times$} in energy efficiency. With mixed-precision quantization, accuracy tracks FP32 within \textbf{$\le$0.9} AP across Deformable and Sparse DETR variants. By converting sparsity into locality—and locality into utilization—\design{} delivers consistent, end-to-end speedups.
\end{abstract}

\begin{comment}
\begin{abstract}
Deformable transformers have emerged as efficient alternatives to dense attention for object detection and multi-modal tasks.
While multi-scale deformable attention (MSDeformAttn) reduces computation by attending to sparse sampling points, hardware performance is bottlenecked by irregular memory access patterns.
Deformable transformers suffer from low arithmetic intensity, bank conflicts, cache misses, and poor batch-level parallelism, confining them in the memory-bound regime.

We present \design, a \emph{cache-optimal, reorder-capable, sparsity-aware accelerator} for end-to-end deformable transformer inference.
\design~comprises three synergistic components: (1) a dedicated MSDeformAttn engine with conflict-free memory access, (2) distance-based out-of-order querying to reorder sparse queries by spatial locality dynamically, and (3) seamless pipelining with heterogeneous GEMM units to facilitate batch-level parallelism across dense and sparse workloads.
Evaluated on both dense and sparse Deformable DETR variants, \design~achieves up to 7.29× speedup and 47.3× energy efficiency over an RTX 4090 GPU.
Compared to state-of-the-art accelerators, \design~delivers 3.26–9.82× higher throughput with full encoder-decoder support.
\end{abstract}
\end{comment}


\vspace{-1mm}
\section{Introduction}\label{sec:intro}
Detection transformers (DETR)~\cite{detr} have attracted widespread attention for their end-to-end object detection capabilities, often outperforming CNN-based detectors in simplicity and adaptability~\cite{detr_var0}.
As a result, DETR-based models have seen rapid adoption in tasks such as 2D recognition and multimodal AI systems~\cite{detr,detr_var0,detr_var1,detr_var2,detr_var3,huang2024tell}.
However, original DETR variants still suffer from high computational cost and slow convergence, mainly due to dense attention mechanisms that obscure gradient flow from input features~\cite{deformable_detr}.

\begin{figure}[tb!]
\centering
\includegraphics[width=0.80\linewidth]{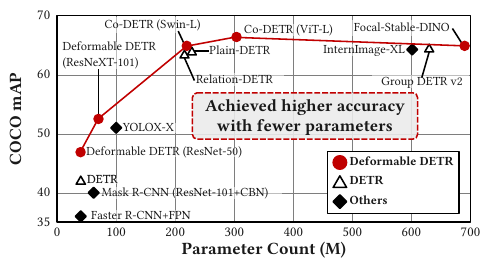}
\vspace{-2mm}
\caption{\textbf{Landscape of SOTA object detection models.}}
\vspace{-4mm}
\label{fig:sota_models}
\end{figure}

Deformable DETR~\cite{deformable_detr} addresses these issues with multi-scale deformable attention (MSDeformAttn), sampling only a handful of spatial locations per query.
Inspired by deformable convolutions~\cite{deformable_conv}, this design significantly reduces FLOPs and accelerates training.
Consequently, as shown in Fig.~\ref{fig:sota_models}, MSDeformAttn underpins state-of-the-art detector modules~\cite{obj_survey, co_detr,dino,dn_detr,sparse_detr,stable_dino,mr_detr} and has been adopted in emerging tasks such as video restoration~\cite{deformable_app0} and large vision--language models~\cite{deformable_app1}.

%%%%%%%%%%%%%%%%%%%%%%%%%%%%%%%%%%%%%%%%%%
%%%%%%%%%%%%%%%%%%%%%%%%%%%%%%%%%%%%%%%%%%

%%%%%%%%%%%%%%%%%%%%%%%%%%%%%%%%%%%%%%%%%%
\begin{figure}[tb!]
\centering
\includegraphics[width=0.85\linewidth]{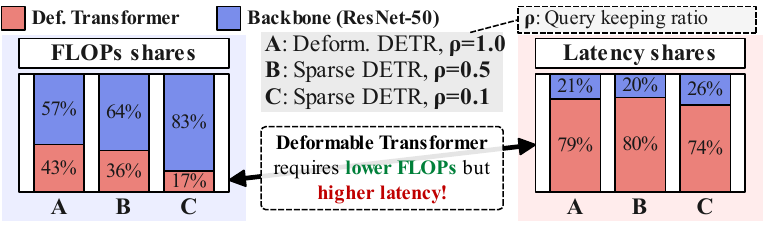}
\vspace{-2mm}
\caption{\textbf{FLOPs vs.\ latency on RTX~4090} comparing the backbone to the deformable transformer block.}
\vspace{-4mm}
\label{fig:fig_flops}
\end{figure}
%%%%%%%%%%%%%%%%%%%%%%%%%%%%%%%%%%%%%%%%%%

%%%%%%%%%%%%%%%%%%%%%%%%%%%%%%%%%%%%%%%%%%
\begin{figure*}[tb!]
\centering
\includegraphics[width=0.9\linewidth]{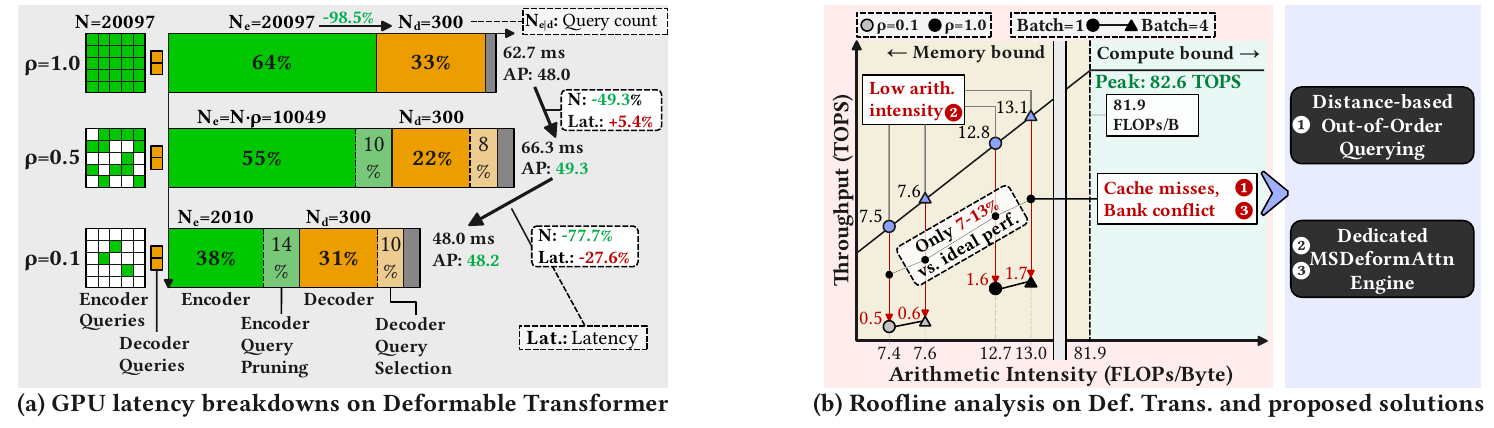}
\vspace{-2mm}
\caption{\textbf{GPU performance breakdowns of Deformable DETR and query-pruned models on RTX~4090}. (\textbf{a}) Latency across internal blocks. (\textbf{b}) Roofline view of bottlenecks and mitigation levers.}
\vspace{-5mm}
\label{fig:motivation}
\end{figure*}
%%%%%%%%%%%%%%%%%%%%%%%%%%%%%%%%%%%%%%%%%%

Despite MSDeformAttn’s algorithmic efficiency, practical deployments on GPUs expose notable latency and throughput drawbacks~\cite{deform_slow0, defa, ueda}, limiting adoption in real systems~\cite{deform_slow1}.
Our profiling on an RTX~4090 highlights four observations:

\textbf{1. Fewer FLOPs, higher latency (Fig.~\ref{fig:fig_flops}).}
Although the backbone accounts for most FLOPs, the \emph{deformable transformer block dominates run time}.
MSDeformAttn’s mathematically sparse pattern \emph{fragments memory access}, triggering frequent stalls.

\textbf{2. Pruning reduces FLOPs but not latency proportionally (Fig.~\ref{fig:motivation}(a)).}
Query pruning~\cite{sparse_transformer} lowers FLOPs roughly with the keeping ratio $\rho$~\cite{sparse_detr}, but latency gains remain modest (e.g., even at $\rho{=}0.1$) due to scattered sampling overheads.

\textbf{3. Decoder cost remains substantial even with 98.5\% fewer queries (Fig.~\ref{fig:motivation}(a), top).}
The decoder uses a tiny set of queries (e.g., 300; a \textbf{98.5\% reduction} vs.\ the encoder’s 20{,}097), yet it still \textbf{accounts for $\sim$33\% of runtime}, indicating an \textbf{order-of-magnitude higher per-query cost}.
Modern Deformable DETR variants exacerbate this by increasing $N_d$ (e.g., $N_d{=}900$~\cite{co_detr,mr_detr}).

\textbf{4. Three fundamental bottlenecks (Fig.~\ref{fig:motivation}(b)).}
A roofline view identifies
cache misses (\circled{1}),
low arithmetic intensity (\circled{2}),
frequent bank conflicts (\circled{3})~\cite{defa}
as dominant causes of \textbf{low PE utilization} on current platforms.

These results reveal a mismatch between MSDeformAttn’s sparse behavior and modern GPU architectures, motivating an \textbf{algorithm--architecture co-design} that preserves model outputs while reorganizing execution for locality and utilization.

Most transformer accelerators target standard attention where accesses are denser and more regular~\cite{ssr,acceltran,hgpipe,random2,random3}.
In contrast, \emph{MSDeformAttn is inherently irregular}, sampling from scattered references across multi-level features.
Recent MSDeformAttn accelerators~\cite{defa,ueda} leave critical gaps:
\textbf{(i)} \textbf{DEFA}~\cite{defa} processes queries in a fixed sequential order and relies on a sliding-window cache, which is \textbf{ill-suited to the decoder and pruned encoder} where sparsity dominates, leading to \textbf{severe cache misses}; moreover, it runs bilinear interpolation, Softmax, and linear layers as \textbf{separate micro-kernel passes}, forcing full memory load--store between passes.
\textbf{(ii)} \textbf{UEDA}~\cite{ueda} \textbf{struggles to scale to standard Deformable DETR} settings (e.g., $N_e{=}20{,}097$, $D{=}256$) due to \textbf{FPGA resource constraints} and limited algorithmic leverage.
To our knowledge, no prior art (1) \textbf{addresses sparse query irregularity at runtime} and (2) \textbf{unifies encoder, decoder, and FFN} in one execution path.

We present \textbf{\design}, an \emph{algorithm--architecture co-design} that turns deformable attention into cache-friendly work while preserving model semantics.
Unlike fixed traversal/sliding-window schemes~\cite{defa} or FPGA-restricted pipelines~\cite{ueda}, \design~introduces a runtime, proximity-aware reordering that restructures execution for locality.
Our contributions are:

\begin{itemize}[leftmargin=*,topsep=2pt,itemsep=2pt]
\item \textbf{DOOQ runtime (query-locality scheduling with aligned prefetch).}
\emph{Distance-based Out-of-Order Querying (DOOQ)} reorders queries by spatial proximity to raise cache hit rate. Its lookup window exposes predictable addresses, enabling \emph{double-buffered} prefetch aligned to the schedule, overlapping fetch with compute and cutting cache-miss stalls (\circled{1}).

\item \textbf{Fused MSDeformAttn execution path.}
A \emph{single-pass} path that fuses sampling, weighting (Softmax), and aggregation to lift arithmetic intensity (\circled{2}) and avoid intermediate stores; conflict-aware buffers curb bank conflicts (\circled{3}), addressing inefficiencies in prior designs~\cite{defa}.

\item \textbf{RTL prototype \& full-stack evaluation.}
A synthesizable RTL of the full accelerator (DOOQ scheduler, feature cache, fused core, gather–scatter, GEMM I/F), functionally verified against a reference MSDeformAttn, with cycle-accurate simulation, post-synthesis PPA, and end-to-end throughput/energy vs.\ RTX~4090 and DEFA (incl.\ ablations on $w_d$, $r$).
\end{itemize}

Across Deformable DETR variants, \design~achieves up to \textbf{7.29$\times$ higher throughput} and \textbf{47.3$\times$ better energy efficiency} than an RTX~4090 GPU. Compared to DEFA~\cite{defa}, \design~improves throughput by \textbf{3.26--9.82$\times$}, attributable to DOOQ’s locality gains together with the fused execution path.

\section{Preliminaries \& Hardware Challenges}\label{sec:prelim}
We give a \textbf{hardware-centric view of MSDeformAttn} and explain why its \emph{sparse, data-dependent} access pattern dominates latency on existing platforms. We then show how common sparsification (e.g., \emph{query pruning}) further degrades locality, motivating our algorithm--architecture co-design.

\vspace{-1mm}
\subsection{MSDeformAttn: A Hardware View}
\begin{figure}[t]
\centering
\vspace{-4mm}
\includegraphics[width=0.8\linewidth]{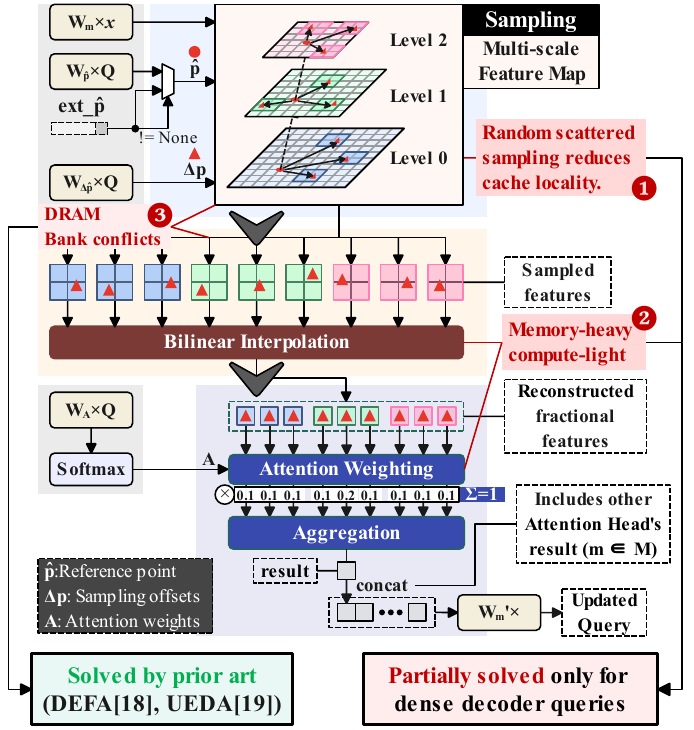}
\vspace{-1mm}
\caption{\textbf{MSDeformAttn dataflow linked to Eq.~\eqref{eq:deform_atten}}, with hardware bottlenecks \circled{1}--\circled{3} marked at the memory touchpoints: \circled{1} cache misses from scattered samples; \circled{2} low arithmetic intensity (few ops per byte); \circled{3} bank conflicts around fractional 2$\times$2 fetches.}
\vspace{-5mm}
\label{fig:deform_atten}
\end{figure}

MSDeformAttn~\cite{deformable_detr} computes, for query $q$, reference point $p$ and spatial embedding feature map $x$,
\begin{equation} \label{eq:deform_atten}
\begin{aligned} 
    MSDeformAttn(q, p, x) = &\\
    \sum_{m=1}^{M} W_m \cdot \Biggl[\sum_{l=1}^L \sum_{k=1}^K A_{qmlk} &\cdot W_m' \cdot x(\hat{p}_q + \Delta p_{qmlk})\Biggr]
\end{aligned}
\end{equation}
where $m,l,k$ index head, feature level, and sampling point. Each query gathers a few \emph{fractional} locations per head across multi-level features (via 2$\times$2 bilinear interpolation), applies attention weights $A$, and aggregates the result. FLOPs are modest, but the \emph{irregular, multi-level gathers} fragment memory access (Fig.~\ref{fig:deform_atten}), yielding \circled{1} cache misses, \circled{2} low arithmetic intensity, and \circled{3} frequent bank conflicts~\cite{defa}.

%%%%%%%%%%%%%%%%%%%%%%%%%%%%%%%%%%%%%%%%%%
%%%%%%%%%%%%%%%%%%%%%%%%%%%%%%%%%%%%%%%%%%

\vspace{-1mm}
\subsection{Query Pruning Hurts Data Locality}
\begin{figure}[t]
\centering
\vspace{-0.5mm}
\includegraphics[width=\linewidth]{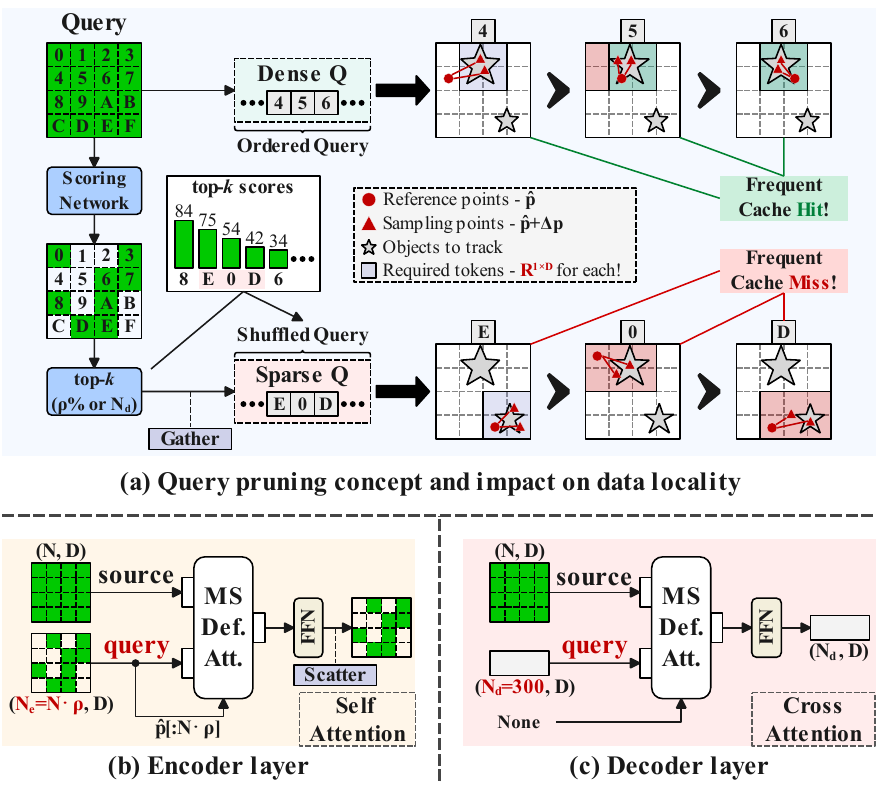}
\vspace{-6mm}
\caption{\textbf{Sparsity amplifies irregular access.} (\textbf{a}) Top-$k$ pruning cuts FLOPs but scatters queries, reducing cache hit rate. (\textbf{b}) Sparse encoder (top-$\rho\%$). (\textbf{c}) Decoder using top-$N_d$ tokens. Both induce less cache-friendly patterns.}
\vspace{-4mm}
\label{fig:sparse_detr}
\end{figure}

As shown in Fig. \ref{fig:sparse_detr}, Sparse DETR~\cite{sparse_detr} prunes encoder queries to the top-$\rho\%$ to reduce FLOPs, and decoder variants select the top-$N_d$ tokens from encoder outputs. These steps \emph{scatter and shuffle} the surviving queries in space, further degrading spatial/temporal locality. As a result, latency on GPUs does not scale with the FLOP reduction~\cite{sparse_transformer, random0, random1}.

%%%%%%%%%%%%%%%%%%%%%%%%%%%%%%%%%%%%%%%%%%
%%%%%%%%%%%%%%%%%%%%%%%%%%%%%%%%%%%%%%%%%%

\vspace{-1mm}
\subsection{Fundamental Bottlenecks and Implications}
Across encoder and decoder (and under pruning), deformable transformers face \textbf{three} hardware bottlenecks:
\begin{itemize}[leftmargin=*,topsep=1pt,itemsep=1pt]
    \item[\circled{1}] \textbf{Cache misses:} scattered, cross-level gathers defeat locality.
    \item[\circled{2}] \textbf{Low arithmetic intensity:} few arithmetic ops per fetched byte leave compute underutilized.
    \item[\circled{3}] \textbf{Bank conflicts:} fractional 2$\times$2 patch fetches elevate collision risk~\cite{defa}.
\end{itemize}
These factors limit throughput and efficiency on extent platforms, even FLOPs are reduced. Closing this gap requires a \emph{runtime} that \textbf{restores locality} and makes next accesses predictable enough to \emph{drive double-buffered (ping--pong) prefetch}, paired with a fused execution path that avoids redundant load--store traffic—principles we adopt in our co-design.

%%%%%%%%%%%%%%%%%%%%%%%%%%%%%%%%%%%%%%%%%%
\begin{figure}[tbh]
\centering
\includegraphics[width=\linewidth]{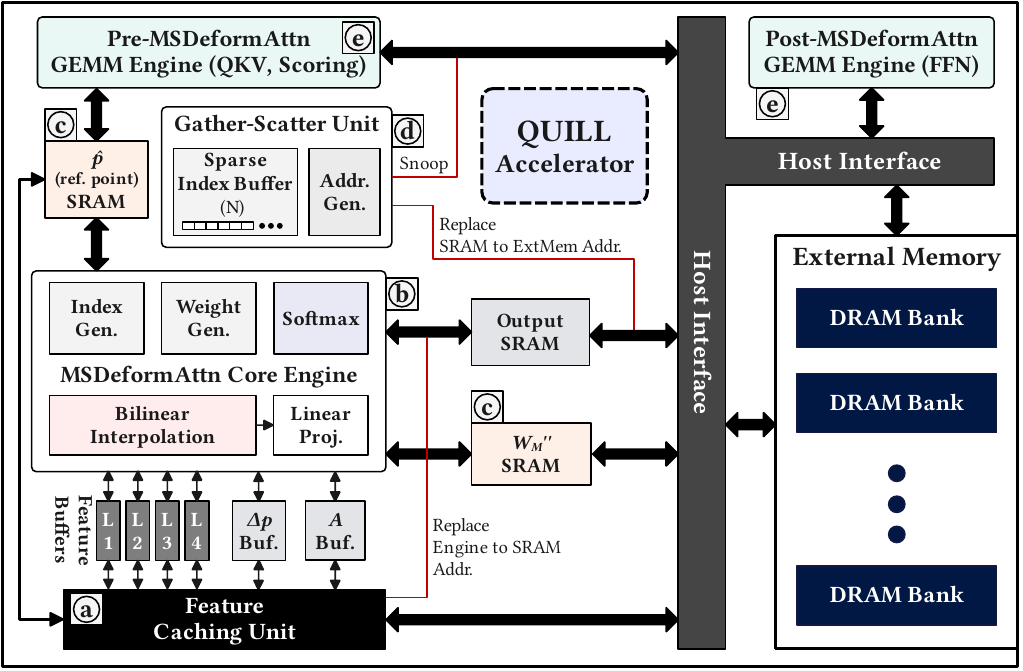}
\vspace{-6mm}
\caption{\textbf{Overview of \design.} The design has two pillars: (\textcircled{a}) a feature caching unit driven by DOOQ and aligned double-buffered prefetch to mitigate cache misses (\circled{1}); and (\textcircled{b}) a fused MSDeformAttn core that executes interpolation, Softmax, aggregation, and projection in one pass to lift arithmetic intensity (\circled{2}) and curb conflict hotspots (\circled{3}). Support blocks include (\textcircled{c}) on-chip reuse SRAMs for small, frequently reused tensors, (\textcircled{d}) a light gather–scatter path for sparse I/O, and (\textcircled{e}) standard GEMM engines for surrounding dense layers. \emph{Core of \design: DOOQ with aligned prefetch, and a fused single-pass core (vs.\ fixed/sliding traversal).}}
\label{fig:top_arch}
\vspace{-3mm}
\end{figure}
%%%%%%%%%%%%%%%%%%%%%%%%%%%%%%%%%%%%%%%%%%

\section{Algorithm--Architecture Co-Design}\label{sec:arch}
\design\ targets the three bottlenecks from Sec.~\ref{sec:prelim} by turning deformable attention into cache-friendly work. At a high level, DOOQ chooses a spatially local execution order for queries; those predicted next addresses drive a region-based prefetch into an alternate cache buffer; the fused core consumes the current buffer and completes interpolation, weighting (Softmax), aggregation, and projection without spilling intermediates; and small, reusable tensors stay on-chip while a light gather–scatter makes sparse I/O contiguous. The surrounding dense layers run on standard systolic GEMMs. Unlike fixed traversal or sliding-window caches~\cite{defa} and FPGA-restricted pipelines~\cite{ueda}, the schedule adapts online and exposes predictable next addresses that a prefetcher can exploit. We refer to this DOOQ→look-ahead→ping–pong path as a schedule-aware prefetch (SAP) loop.

Our co-design follows two principles. First, use a \emph{runtime, content-agnostic schedule} (DOOQ) that restores locality without changing model semantics and exposes predictable next addresses for prefetch. Second, execute MSDeformAttn in a \emph{single pass} so locality gains translate directly into higher arithmetic intensity and fewer stalls.

\subsection{(\textcircled{a}) Feature Caching Unit: DOOQ with aligned prefetch}
\begin{figure}[t]
\centering
\includegraphics[width=\linewidth]{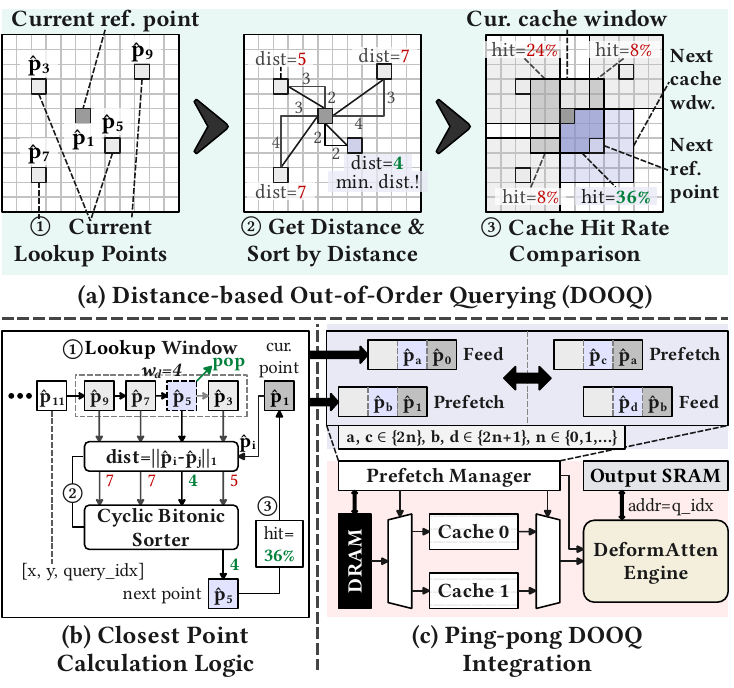}
\vspace{-7mm}
\caption{\textbf{DOOQ and aligned prefetch.} DOOQ chooses spatially closest queries within a window; a lightweight distance+sorting (cyclic bitonic) unit selects the next set; and a ping–pong prefetch overlaps memory with compute using addresses from the DOOQ window. Together these form the schedule-aware prefetch (SAP) loop: schedule $\rightarrow$ look-ahead $\rightarrow$ overlapped prefetch.}
\label{fig:dooq}
\vspace{-5mm}
\end{figure}

The feature cache (Fig.~\ref{fig:top_arch}, \textcircled{a}) forms a closed loop: select the next queries, prefetch their regions, and compute on the current buffer. DOOQ supplies the order; a small selector realizes it; and double buffering turns predictable addresses into overlap between memory and compute. The loop is illustrated in Fig.~\ref{fig:dooq}, and summarized in Alg.~\ref{alg:dooq}.

Within a lookup window of size $w_d$, DOOQ prefers the next queries whose reference points $\hat p$ are spatially closest to the current one, so adjacent queries reuse many of the same feature lines across levels. If $M(q)$ is the set of cache lines needed by query $q$, miss-driven stall for execution order $\sigma$ is
\begin{equation}
T_{\text{stall}}(\sigma)=\sum_{i=1}^{n-1} t_{\text{fetch}}\cdot\big|M(q_{\sigma(i+1)})\setminus M(q_{\sigma(i)})\big| ,
\label{eq:tstall}
\end{equation}
and DOOQ acts directly on this set difference. Algorithm~\ref{alg:dooq} (line~2) performs the greedy nearest-neighbor selection in $\ell_1$; Fig.~\ref{fig:dooq} depicts the corresponding distance+sorting unit. Unlike fixed traversal or sliding-window caches~\cite{defa}, the order adapts online to the current distribution of $\hat p$ while keeping hardware simple (we use $\ell_1$ to avoid squaring and square roots with negligible impact on reuse).

Candidates pass through a cyclic bitonic sorter (Fig.~\ref{fig:dooq}, “distance+sorting”) that reuses compare-and-swap elements over multiple cycles; the per-query slack (about $D/p_d$ cycles) amortizes its cost for typical $w_d$. Area scales linearly with $w_d$ (time-multiplexed $O(w_d)$ comparators over $O(\log^2 w_d)$ steps), and timing fits within the $D/p_d$ slack. This guarantees a one-step look-ahead: when the core begins $q_{\sigma(i)}$, the cache already knows the reference points for $q_{\sigma(i+1)}$.

%%%%%%%%%%%%%%%%%%%%%%%%%%%%%%%%%%%%%%%%%%
%%%%%%%%%%%%%%%%%%%%%%%%%%%%%%%%%%%%%%%%%%

That look-ahead drives supply. The prefetch in Alg.~\ref{alg:dooq} (line~3) issues a regional fetch around each chosen $\hat p$ into the alternate buffer while the core consumes the current one (the ping–pong path in Fig.~\ref{fig:dooq}). At the handoff, only incremental lines are fetched if the next query stays within the prefetched region. Under ping–pong, the residual stall per step is $\max\{0,\,t_{\text{fetch}}(q_{\sigma(i+1)}\,|\,w_d)-t_{\text{comp}}(q_{\sigma(i)})\}$, and DOOQ shrinks the first term by maximizing the overlap between $M(q_{\sigma(i)})$ and $M(q_{\sigma(i+1)})$. With average hit rate $h$ and regional reuse $\rho$, the covered miss time per step is approximately $\min\{t_{\text{comp}},\,(1-h\rho)\,t_{\text{fetch}}\}$.

Flow control is elastic: the emit/update in Alg.~\ref{alg:dooq} (line~4) advances the loop; FIFOs handle backpressure, a small per-level victim path recovers rare off-region offsets, and query IDs preserve semantic order.

%%%%%%%%%%%%%%%%%%%%%%%%%%%%%%%%%%%%%%%%%%
\vspace{-2mm}
\begin{algorithm}[h]
\caption{DOOQ (schedule-aware prefetch loop)}
\label{alg:dooq}
\begin{algorithmic}[1]
\REQUIRE lookup window $\mathcal{W}$ of reference points $\{\hat p\}$, current query $q_i$, region radius $r$
\WHILE{$\mathcal{W}\neq\emptyset$}
  \STATE $q^\star \gets \arg\min_{q\in\mathcal{W}} \|\hat p_q-\hat p_{q_i}\|_1$ \hfill{\scriptsize selection}
  \STATE \textbf{PrefetchRegion}$(\hat p_{q^\star}, r)$ \hfill{\scriptsize to alternate buffer}
  \STATE \textbf{Emit}$(q^\star)$; \quad $\mathcal{W}\gets\mathcal{W}\setminus\{q^\star\}$; \quad $q_i\gets q^\star$ \hfill{\scriptsize update}
\ENDWHILE
\end{algorithmic}
\end{algorithm}
%%%%%%%%%%%%%%%%%%%%%%%%%%%%%%%%%%%%%%%%%%
\vspace{-5mm}

\subsection{(\textcircled{b}) Fused MSDeformAttn Core}
The core turns locality into throughput by avoiding intermediate spills; we execute MSDeformAttn in a single pass. Folding the projections into $W_m''{=}W_m\!\cdot\!W_m'$ yields
\begin{equation} \label{eq:deform_atten_transformed}
\begin{aligned} 
    MSDeformAttn(q, p, x) = &\\
    \sum_{m=1}^{M} W_m'' \cdot \Biggl[\sum_{l=1}^L \sum_{k=1}^K A_{qmlk} &\cdot x(\hat{p}_q + \Delta p_{qmlk})\Biggr]
\end{aligned}
\end{equation}
so interpolation, Softmax, aggregation, and projection proceed without intermediate writes that depress arithmetic intensity. Index and weight generators produce 2$\times$2 coordinates and bilinear weights $w_{ij}$; the same $w_{ij}$ broadcasts across the $D$-wide vector, which keeps the multiplier count low. The 2$\times$2 gather plus interpolation uses four multipliers and three adders per element, with the small generators shared across samples over multiple cycles. Softmax runs in-core to avoid traffic, using shared exp/add hardware (Padé-based exponential) across cycles. Aggregation feeds a projector backed by on-chip storage for $W_m''$, which is loaded once per block and reused across queries. Level buffers are laid out so the common 2$\times$2 access pattern maps to distinct banks; the DOOQ schedule further reduces simultaneous contention across levels, trimming conflict hotspots without requiring large SRAMs. We reorder across \emph{queries} only; within each query the $L{\times}K$ sampling order is unchanged, and Softmax is computed per query, preserving semantics.

%%%%%%%%%%%%%%%%%%%%%%%%%%%%%%%%%%%%%%%%%%
\begin{figure}[t]
\centering
\includegraphics[width=\linewidth]{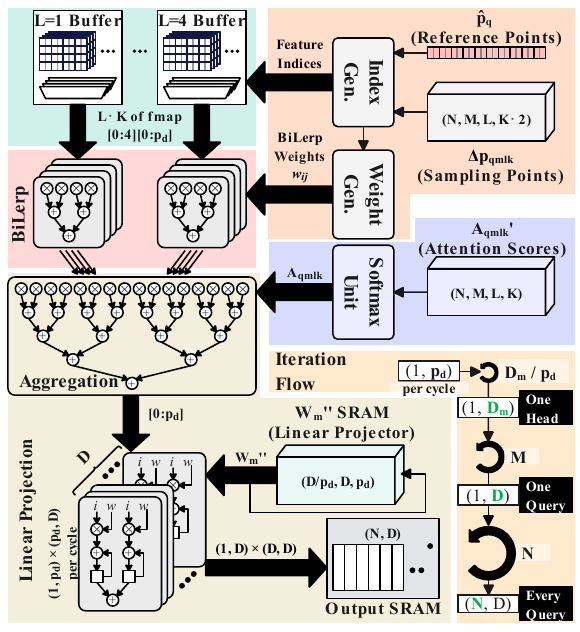}
\vspace{-5mm}
\caption{\textbf{Fused MSDeformAttn core.} Interpolation, Softmax, aggregation, and projection ($W_m''$) are executed in one pass; level buffers are organized to serve 2$\times$2 fetches with minimal collisions.}
\label{fig:engine}
\vspace{-5mm}
\end{figure}
%%%%%%%%%%%%%%%%%%%%%%%%%%%%%%%%%%%%%%%%%%

\subsection{(\textcircled{c}–\textcircled{e}) Support blocks}
Small, frequently reused tensors stay on-chip to cut redundant transfers and help arithmetic intensity: reference points $\hat p$ (two coordinates per query) and the projector $W_m''$ (sized by $M$ and $D$). If $D$ or $M$ grows, $W_m''$ streams in stripes along $D$ with a tiny double buffer. For sparse encoders (top-$\rho\%$) and token-selection decoders (top-$N_d$), a light gather–scatter unit remaps indices so external memory accesses remain burst-friendly; results carry query IDs so downstream order is preserved. The surrounding dense layers (pre-attention/projection and FFN) run on standard systolic GEMM engines.

In summary, DOOQ raises locality so fewer new lines are fetched; aligned prefetch hides most remaining latency; and the fused pass converts the locality into sustained utilization and end-to-end speedups. Section~\ref{sec:eval} quantifies how $w_d$ and $r$ affect hit rate, stall time, and throughput across encoder/decoder and pruned settings, and attributes gains to improved hit rate $h$, regional reuse $\rho$, and fused-pass intensity.

%%%%%%%%%%%%%%%%%%%%%%%%%%%%%%%%%%%%%%%%%%%%%%%%%%%%%%%%%%%%%%%%%%%%%%%%%%%
\section{Evaluation}\label{sec:eval}
We evaluate \design{} end-to-end and quantify how its schedule-aware prefetch and fused single-pass execution translate into system gains.
We implement a synthesizable RTL generator in Chisel~\cite{chisel}, verify with Verilator~5.009, and synthesize with Synopsys Design Compiler using a topographical flow~\cite{topo} in TSMC 28\,nm HVT at 1\,GHz. External memory is modeled as HBM2 (256\,GB/s, 1.21\,pJ/bit~\cite{hbm2}). Unless noted, we use COCO 2017~\cite{coco} and standard Deformable DETR settings $\big(N_e{=}\lceil 20097\!\cdot\!\rho\rceil,\; N_d{=}300,\; D{=}256,\; M{=}8,\; L{=}4,\; K{=}4\big)$, consistent with common SOTA variants~\cite{co_detr,dino,dn_detr,sparse_detr,stable_dino,mr_detr}. The region-prefetch radius $r$ is \emph{fixed} per level to the empirical maximum sampling offset from the model; we ablate only the DOOQ window $w_d$.

%%%%%%%%%%%%%%%%%%%%%%%%%%%%%%%%%%%%%%%%%%%%%%%%%%%%%%%%%%%%%%%%%%%%%%%%%%%%%%%%%%%%%%%%%%%%%%%%%%%%%%%%%%%%%%%%%%%%%%%%%%%%%%
\subsection{Ablation: DOOQ Lookup Window ($w_d$)}
\begin{figure}[tb!]
\centering
\includegraphics[width=\linewidth]{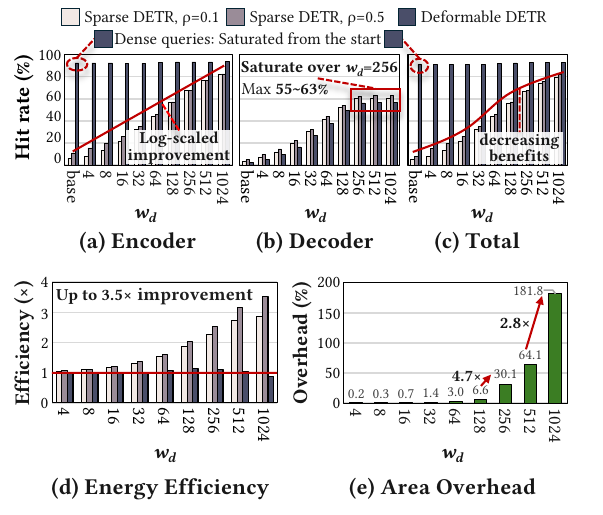}
\vspace{-7mm}
\caption{\textbf{DOOQ sweep over $w_d$.} Encoder/decoder/total hit rate, memory energy efficiency, and area vs.\ a same-capacity direct-mapped cache.}
\label{fig:graph_dooq_cache_hit}
\vspace{-3mm}
\end{figure}

Fig.~\ref{fig:graph_dooq_cache_hit} sweeps $w_d$ against a same-capacity direct-mapped baseline (no DOOQ logic) and reports encoder/decoder/total hit rate, memory energy efficiency, and area.

Encoder locality grows quickly with $w_d$ (Fig.~\ref{fig:graph_dooq_cache_hit}(a)) and exceeds \textbf{80\%} by $w_d{=}1024$—a \textbf{13.9$\times$} improvement over baseline—because spatially adjacent queries reuse feature lines across levels once the window exposes them to the cache. Decoder locality (Fig.~\ref{fig:graph_dooq_cache_hit}(b)) saturates at \textbf{55–63\%} near $w_d{\approx}256$: with $N_d{=}300$ candidates, the window already spans most near neighbors, and the residual accesses are inherently scattered by the token-selection policy. Total hit rate (Fig.~\ref{fig:graph_dooq_cache_hit}(c) tracks the encoder trend as the encoder dominates the query count.

Energy improves monotonically with $w_d$ (Fig.~\ref{fig:graph_dooq_cache_hit}(d)) as fewer new lines are fetched per step in Equation~\eqref{eq:tstall}, while the time-multiplexed sorter keeps area growth predictable (Fig.~\ref{fig:graph_dooq_cache_hit}(e)): $O(w_d)$ comparators reused over $O(\log^2 w_d)$ steps fit in the per-query slack ($D/p_d$ cycles) without timing risk. Balancing hit, energy, and area, we choose \textbf{$w_d{=}512$} for the rest of the study: up to \textbf{10.0$\times$} hit-rate gain vs.\ baseline, \textbf{1.2–3.1$\times$} memory-energy improvement, and $<\textbf{0.42\,mm}^2$ overhead—an attractive energy return per mm$^2$.

%%%%%%%%%%%%%%%%%%%%%%%%%%%%%%%%%%%%%%%%%%
%%%%%%%%%%%%%%%%%%%%%%%%%%%%%%%%%%%%%%%%%%

%%%%%%%%%%%%%%%%%%%%%%%%%%%%%%%%%%%%%%%%%%%%%%%%%%%%%%%%%%%%%%%%%%%%%%%%%%%%%%%%%%%%%%%%%%%%%%%%%%%%%%%%%%%%%%%%%%%%%%%%%%%%%%
\subsection{End-to-End Latency and Throughput}
\begin{figure}[tb!]
\centering
\includegraphics[width=\linewidth]{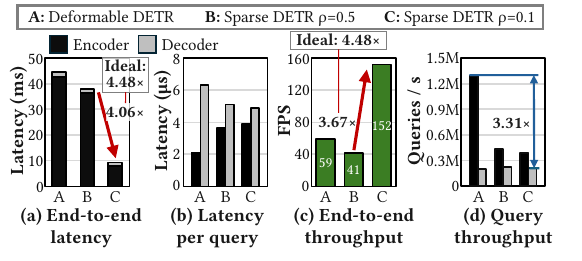}
\vspace{-6mm}
\caption{\textbf{End-to-end performance.} (\textbf{a}) Overall latency. (\textbf{b}) Per-query latency. (\textbf{c}) Overall throughput. (\textbf{d}) Query throughput. Sparsity scales near-ideally ($\rho{=}0.5\!\rightarrow\!0.1$).}
\label{fig:graph_perf}
\vspace{-5mm}
\end{figure}

Fig.~\ref{fig:graph_perf} shows how locality gains turn into system-level speedups.
Moving from $\rho{=}0.5$ to $\rho{=}0.1$ reduces end-to-end latency by \textbf{4.06$\times$} (Fig.~\ref{fig:graph_perf}(a)), close to the ideal \textbf{4.48$\times$} dictated by query counts (10{,}349$\!\rightarrow$2{,}310), and clearly exceeding the GPU’s scaling baseline (cf.\ Fig.~\ref{fig:motivation}). The small gap at $\rho{=}0.5$ versus dense is consistent with the mid-sparsity hit-rate dip: encoder requests still reuse well, while decoder misses begin to dominate.

Per-query behavior (Fig.~\ref{fig:graph_perf}(b)) matches this: dense encoder queries are fastest; sparsity tightens decoder locality and lowers median per-query latency, yet some off-region accesses remain that the local reordering in Equation~\eqref{eq:tstall} cannot fully hide—explaining the residual gap to ideal scaling.

Throughput scales accordingly. Overall throughput (Fig.~\ref{fig:graph_perf}(c)) reaches \textbf{3.67$\times$} at $\rho{=}0.1$ vs.\ $\rho{=}0.5$, near the ideal \textbf{4.48$\times$}. Query throughput (Fig.~\ref{fig:graph_perf}(d)) is highest for the dense encoder—up to \textbf{3.31$\times$} over sparse decoder queries—thanks to stronger reuse and more uniform access patterns. In practice, aggressive pruning ($\rho{\rightarrow}0.1$) delivers not only fewer FLOPs but also sustained wall-clock speedups.

%%%%%%%%%%%%%%%%%%%%%%%%%%%%%%%%%%%%%%%%%%
%%%%%%%%%%%%%%%%%%%%%%%%%%%%%%%%%%%%%%%%%%

%%%%%%%%%%%%%%%%%%%%%%%%%%%%%%%%%%%%%%%%%%%%%%%%%%%%%%%%%%%%%%%%%%%%%%%%%%%%%%%%%%%%%%%%%%%%%%%%%%%%%%%%%%%%%%%%%%%%%%%%%%%%%%
\subsection{Area and Power Breakdown}
We next quantify the silicon budget of \design. Table~\ref{tab:area_power} maps blocks to Fig.~\ref{fig:top_arch}; gather–scatter (\textcircled{d}) is folded into \emph{Others}.

\begin{table}[t]
\centering
\small
\caption{Hardware evaluation at 1\,GHz, 0.81\,V (TSMC 28\,nm HVT). Markers (\textcircled{a}–\textcircled{e}) follow Fig.~\ref{fig:top_arch}. “Others” includes gather–scatter (\textcircled{d}) and clk/host interface.}
\label{tab:area_power}
\vspace{-2mm}
\resizebox{\linewidth}{!}{
\begin{tabular}{lrrrr}
\hline
\textbf{Block} & \textbf{Area} & \textbf{Area} & \textbf{Power} & \textbf{Power} \\
 & (mm$^2$) & (\%) & (mW) & (\%) \\
\hline
DOOQ scheduler/sorter (\textcircled{a})         & 0.416 & 3.83  & 83.4   & 2.92 \\
Feature cache L1–L4 (\textcircled{a})           & 1.306 & 12.04 & 241.6  & 8.44 \\
Fused MSDeformAttn core (\textcircled{b})       & 0.146 & 1.35  & 182.7  & 6.39 \\
GEMM engines (32$\times$32+64$\times$64) (\textcircled{e}) & 1.919 & 17.69 & 1406.0 & 49.15 \\
On-chip SRAMs (other) (\textcircled{c})         & 6.297 & 58.04 & 788.6  & 27.56 \\
Others (incl.\ \textcircled{d})                 & 0.766 & 7.06  & 158.6  & 5.54 \\
\hline
\textbf{Total}                                  & \textbf{10.85} & \textbf{100} & \textbf{2860.9} & \textbf{100} \\
\hline
\end{tabular}
}
\vspace{-4mm}
\end{table}

Area is memory-centric: on-chip SRAMs (\textcircled{c}) plus the feature cache (\textcircled{a}) occupy \textbf{7.60\,mm$^2$} ($\approx$\textbf{70\%}). Power is compute-centric: the two GEMMs (\textcircled{e}) draw \textbf{49\%}, reflecting FFN/projections. The fused MSDeformAttn core (\textcircled{b}) is compact (1.35\% area, 6.39\% power), while DOOQ control (scheduler/sorter) is small—\textbf{3.83\%} area and \textbf{2.92\%} power—so most of the ``DOOQ budget'' is the cache capacity that enables the hit-rate gains in Fig.~\ref{fig:graph_dooq_cache_hit}. For smaller footprints, first right-size the output SRAM (tiling/streaming), then tune $w_d$; $r$ is fixed by the model’s offset envelope, and the selector scales near-linearly, fitting in the $D/p_d$ slack. With $w_d{=}512$, the cache path adds \textbf{${<}0.42$\,mm$^2$} yet yields up to \textbf{10$\times$} higher hit rate and \textbf{$1.2$--$3.1\times$} memory-energy improvement.

%%%%%%%%%%%%%%%%%%%%%%%%%%%%%%%%%%%%%%%%%%%%%%%%%%%%%%%%%%%%%%%%%%%%%%%%%%%%%%%%%%%%%%%%%%%%%%%%%%%%%%%%%%%%%%%%%%%%%%%%%%%%%%
\subsection{Model Accuracy (Baseline vs.\ QUILL)}
\begin{figure}[tb!]
\centering
\includegraphics[width=\linewidth]{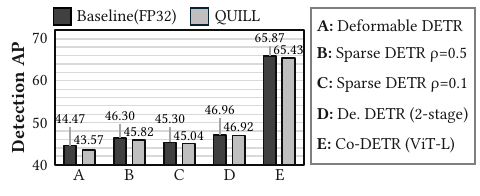}
\vspace{-8mm}
\caption{\textbf{Detection accuracy (AP): FP32 vs.\ QUILL.} A: Deformable DETR; B: Sparse DETR ($\rho{=}0.5$); C: Sparse DETR ($\rho{=}0.1$); D: 2-stage; E: Co-DETR (ViT-L). QUILL is within ${\le}0.9$ AP of FP32.}
\label{fig:graph_acc}
\vspace{-5mm}
\end{figure}

We evaluate AP under FP32 and under QUILL—our fused single-pass fixed-point path with \emph{mixed precision} (W8A8$\!\rightarrow$A18 for linear/projection and bilinear, W16A8$\!\rightarrow$A28 for attention-weighted aggregation/Softmax). Model semantics are unchanged except DOOQ scheduling. Figure~\ref{fig:graph_acc} covers Deformable DETR, Sparse DETR at $\rho{=}0.5$/$0.1$, a two-stage variant, and large Co-DETR (ViT-L). Across all cases, QUILL is within \textbf{0.04–0.90} AP of FP32, preserving accuracy while enabling the locality-driven speedups above.

%%%%%%%%%%%%%%%%%%%%%%%%%%%%%%%%%%%%%%%%%%
%%%%%%%%%%%%%%%%%%%%%%%%%%%%%%%%%%%%%%%%%%

%%%%%%%%%%%%%%%%%%%%%%%%%%%%%%%%%%%%%%%%%%%%%%%%%%%%%%%%%%%%%%%%%%%%%%%%%%%%%%%%%%%%%%%%%%%%%%%%%%%%%%%%%%%%%%%%%%%%%%%%%%%%%%
\subsection{Cross-Platform Analyses}
We benchmark \design{} against a high-end GPU and recent MSDeformAttn accelerators.

\begin{figure}[tb!]
\centering
\includegraphics[width=\linewidth]{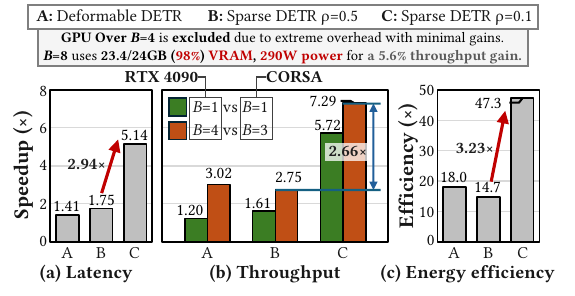}
\vspace{-7mm}
\caption{\textbf{GPU comparison (RTX 4090).} (\textbf{a}) Latency. (\textbf{b}) Throughput vs.\ batch size. (\textbf{c}) $B{=}1$ energy efficiency.}
\label{fig:graph_comp_gpu}
\vspace{-1mm}
\end{figure}

\textbf{vs. RTX 4090.}
Fig.~\ref{fig:graph_comp_gpu} compares latency, throughput vs.\ batch size ($B$), and $B{=}1$ energy. \design{} delivers up to \textbf{5.14$\times$} lower latency, \textbf{7.29$\times$} higher throughput, and \textbf{47.3$\times$} better $B{=}1$ energy efficiency, with the largest margins at $\rho{=}0.1$. Batch scaling is favorable: we exceed the GPU at $B{=}3$ (GPU needs $B{=}4$), avoiding large-$B$ VRAM/power overheads.

%%%%%%%%%%%%%%%%%%%%%%%%%%%%%%%%%%%%%%%%%%
%%%%%%%%%%%%%%%%%%%%%%%%%%%%%%%%%%%%%%%%%%

\textbf{vs. Dedicated accelerators.}
Table~\ref{tab:comp_sota} places \design{} alongside DEFA~\cite{defa} and UEDA~\cite{ueda}. Unlike DEFA (encoder-only, micro-kernelized, INT12) and UEDA (FPGA, reduced MSDA), \design{} accelerates \emph{MSDeformAttn end-to-end with FFN} in a single fused pass while preserving FP32 semantics. This scope plus DOOQ-driven locality yields the \textbf{highest peak throughput} (12.06\,TOPS) and \textbf{area efficiency} (1.11\,TOPS/mm$^2$) with competitive \textbf{power efficiency} (4.23\,TOPS/W). On normalized end-to-end metrics, \design{} sustains \textbf{31.76–95.82 FPS/TOPS} and \textbf{134.3–405.1 FPS/W} (Deformable DETR $\rightarrow$ Sparse DETR, $\rho{=}0.1$). Versus DEFA, throughput utilization and energy efficiency are \textbf{3.26–9.82$\times$} and \textbf{2.01–6.07$\times$} higher; UEDA trails due to FPGA frequency/resource limits and reduced configs.

\begin{table}[t]
\centering
\caption{\textbf{Comparison with other accelerators}, including not only SOTA MSDeformAttn (MSDA) accelerators.}
\vspace{-2mm}
\resizebox{\linewidth}{!}{%
\begin{tabular}{l|c||c|c|c}
\toprule
Metric & Unit
& \begin{tabular}[c]{@{}c@{}}\textbf{This work}\\\textbf{\design}\end{tabular}
& \begin{tabular}[c]{@{}c@{}}\textbf{DAC'24}\\~~~~\textbf{DEFA}\cite{defa}~~~~\end{tabular}
& \begin{tabular}[c]{@{}c@{}}~~~\textbf{ASPDAC'25}~~~\\\textbf{UEDA}\cite{ueda}\end{tabular}
\\ \hline \hline

Workload & -
& \begin{tabular}[c]{@{}c@{}}MSDA, FFN\end{tabular}
& \begin{tabular}[c]{@{}c@{}}MSDA \\ (Encoder only)\end{tabular}           
& \begin{tabular}[c]{@{}c@{}}MSDA \\ (Shrunken$^\mathrm{*}$)\end{tabular}                                                         
\\ \hline

Precision & -
& \begin{tabular}[c]{@{}c@{}}W8A8$\rightarrow$A18,\\W16A8$\rightarrow$A28\end{tabular}                     
& INT12                                                                      
& \begin{tabular}[c]{@{}l@{}}Not specified\end{tabular}                       
\\ \hline \hline

Technology & -
& 28 nm                         
& 40 nm                                 
& FPGA                                                                     
\\ \hline

Area & mm²
& 10.85                                 
& 2.63
& -
\\ \hline

Power & mW
& 2860.9
& 99.8
& 3070.0
\\ \hline

Frequency & MHz
& 1000        
& 400
& 200
\\ \hline

Peak Throughput & TOPS
& \textbf{12.06}
& 0.42
& 0.68
\\ \hline \hline

Power Efficiency & TOPS / W
& \textbf{4.23}
& 4.19
& 0.22
\\ \hline

Area Efficiency &~TOPS / mm²~
& \textbf{1.11}
& 0.16
& -
\\ \hline \hline

\begin{tabular}[c]{@{}l@{}}Normalized$^\mathrm{*}$ \\ Power Efficiency\end{tabular} & TOPS / W
& \textbf{4.23}
& \textbf{6.84}
& -
\\ \hline

\begin{tabular}[c]{@{}l@{}}Normalized$^\mathrm{*}$ \\ Area Efficiency\end{tabular} & TOPS / mm²
& \textbf{1.11}
& 0.28
& -
\\ \hline \hline 

\begin{tabular}[c]{@{}l@{}}Normalized$^\mathrm{*}$\\Throughput\end{tabular} & FPS
& \textbf{1,270.5$^\mathrm{a}$-3832.7}$^\mathrm{b}$
& 390.3
& 108.0
\\ \hline

\begin{tabular}[c]{@{}l@{}}Utilization$^\mathrm{*}$\\Efficiency\end{tabular} & FPS / TOPS
& \textbf{31.76$^\mathrm{a}$-95.82$^\mathrm{b}$}
& 9.76
& 2.70
\\ \hline

\begin{tabular}[c]{@{}l@{}}Normalized$^\mathrm{*}$\\Energy Efficiency\end{tabular} & FPS / W
& \textbf{134.3$^\mathrm{a}$-405.1$^\mathrm{b}$}
& 66.7
& -
\\ \bottomrule
\multicolumn{5}{l}{\scriptsize $^\mathrm{a}$Deformable DETR. $^\mathrm{b}$Sparse DETR $\rho$=0.1. $^\mathrm{*}$Normalized to 28 nm process using the industry scaling model in \cite{ppa_scaling}.} \\
%\multicolumn{5}{l}{\scriptsize $^\mathrm{**}$Normalized by matching peak TOPS with RTX 3090Ti using the method in \cite{defa, elsa, dota}.} \\
\end{tabular}}
\vspace{-6mm}
\label{tab:comp_sota}
\end{table}

\noindent Overall, converting deformable attention into cache-friendly, single-pass work lets \design{} translate sparsity into end-to-end gains on real models while staying within ${\le}0.9$\,AP of FP32.
%%%%%%%%%%%%%%%%%%%%%%%%%%%%%%%%%%%%%%%%%%%%%%%%%%%%%%%%%%%%%%%%%%%%%%%%%%%
\section{Conclusion}\label{sec:conclusion}
We presented \design, an algorithm–architecture co-design that turns MSDeformAttn into cache-friendly, single-pass work. By coupling DOOQ’s scheduling with a fused MSDeformAttn core and integrated GEMMs, \design delivers near-linear sparse scaling and end-to-end gains: up to 7.29$\times$ higher throughput and 47.3$\times$ better $B{=}1$ energy efficiency than an RTX~4090, and up to 9.82$\times$/6.07$\times$ higher throughput/energy efficiency than prior accelerators.

Three directions remain: (i) adaptive scheduling/prefetch (learned, multi-step) that tunes $w_d$ and cache policy online; (ii) applying the schedule-aware prefetch loop to other sparse ops (cross-attention, video, point clouds) and to larger backbones via streamed $W_m''$ tiling; and (iii) a compiler/runtime that auto-tunes micro-architectural knobs (parallelism $p_d$, GEMM tiling, cache partitioning) with tighter quantization-aware training (QAT) for mixed precision. Together, these extend \design’s core idea—turn sparsity into locality, then utilization—to the next wave of detection and multimodal transformers.

\bibliographystyle{IEEEtran}
\bibliography{IEEEabrv, References/intro, References/rw_hw_arch, References/technical_contents}

\begin{thebibliography}{10}
\providecommand{\url}[1]{#1}
\csname url@samestyle\endcsname
\providecommand{\newblock}{\relax}
\providecommand{\bibinfo}[2]{#2}
\providecommand{\BIBentrySTDinterwordspacing}{\spaceskip=0pt\relax}
\providecommand{\BIBentryALTinterwordstretchfactor}{4}
\providecommand{\BIBentryALTinterwordspacing}{\spaceskip=\fontdimen2\font plus
\BIBentryALTinterwordstretchfactor\fontdimen3\font minus \fontdimen4\font\relax}
\providecommand{\BIBforeignlanguage}[2]{{%
\expandafter\ifx\csname l@#1\endcsname\relax
\typeout{** WARNING: IEEEtran.bst: No hyphenation pattern has been}%
\typeout{** loaded for the language `#1'. Using the pattern for}%
\typeout{** the default language instead.}%
\else
\language=\csname l@#1\endcsname
\fi
#2}}
\providecommand{\BIBdecl}{\relax}
\BIBdecl

\bibitem{detr}
N.~Carion, F.~Massa, G.~Synnaeve, N.~Usunier, A.~Kirillov, and S.~Zagoruyko, ``End-to-end object detection with transformers,'' in \emph{16th Eur. Conf. Comput. Vis. (ECCV)}, Glasgow, United Kingdom, Aug. 2020, p. 213–229.

\bibitem{detr_var0}
Y.~Zhao \emph{et~al.}, ``{DETRs Beat YOLOs on Real-time Object Detection},'' in \emph{IEEE/CVF Conf. Comput. Vis. Pattern Recognit. (CVPR)}, Seattle, WA, USA, Jun. 2024, pp. 16\,965--16\,974.

\bibitem{detr_var1}
X.~Gu, H.~Fan, Y.~Huang, T.~Luo, and L.~Zhang, ``{Context-Guided Spatio-Temporal Video Grounding},'' in \emph{IEEE/CVF Conf. Comput. Vis. Pattern Recognit. (CVPR)}, Seattle, WA, USA, Jun. 2024, pp. 18\,330--18\,339.

\bibitem{detr_var2}
A.~Kamath, M.~Singh, Y.~LeCun, G.~Synnaeve, I.~Misra, and N.~Carion, ``{MDETR - Modulated Detection for End-to-End Multi-Modal Understanding},'' in \emph{IEEE/CVF Int. Conf. Comput. Vis. (ICCV)}, Montreal, QC, Canada, Oct. 2021, pp. 1760--1770.

\bibitem{detr_var3}
J.~Xu \emph{et~al.}, ``{Pixel Aligned Language Models},'' in \emph{IEEE/CVF Conf. Comput. Vis. Pattern Recognit. (CVPR)}, Seattle, WA, USA, Jun. 2024, pp. 13\,030--13\,039.

\bibitem{huang2024tell}
\BIBentryALTinterwordspacing
W.~Huang \emph{et~al.}, ``{Tell Me What to Track: Infusing Robust Language Guidance for Enhanced Referring Multi-Object Tracking},'' \emph{arXiv preprint arXiv:2412.12561}, 2024. [Online]. Available: \url{https://arxiv.org/abs/2412.12561}
\BIBentrySTDinterwordspacing

\bibitem{deformable_detr}
X.~Zhu, W.~Su, L.~Lu, B.~Li, X.~Wang, and J.~Dai, ``{Deformable DETR: Deformable Transformers for End-to-End Object Detection},'' in \emph{Int. Conf. Learn. Represent. (ICLR)}, Virtual, May 2021, pp. 1--16.

\bibitem{deformable_conv}
X.~Zhu, H.~Hu, S.~Lin, and J.~Dai, ``{Deformable ConvNets V2: More Deformable, Better Results},'' in \emph{IEEE/CVF Conf. Comput. Vis. Pattern Recognit. (CVPR)}, Long Beach, CA, USA, Jun. 2019, pp. 9300--9308.

\bibitem{obj_survey}
Z.~Zou, K.~Chen, Z.~Shi, Y.~Guo, and J.~Ye, ``{Object Detection in 20 Years: A Survey},'' \emph{Proc. {IEEE}}, vol. 111, no.~3, pp. 257--276, 2023.

\bibitem{co_detr}
Z.~Zong, G.~Song, and Y.~Liu, ``{DETRs with Collaborative Hybrid Assignments Training},'' in \emph{IEEE/CVF Int. Conf. Comput. Vis. (ICCV)}, Paris, France, Oct. 2023, pp. 6748--6758.

\bibitem{dino}
H.~Zhang \emph{et~al.}, ``{DINO: DETR with Improved DeNoising Anchor Boxes for End-to-End Object Detection},'' in \emph{Int. Conf. Learn. Represent. (ICLR)}, Kigali, Rwanda, May 2023, pp. 1--19.

\bibitem{dn_detr}
F.~Li, H.~Zhang, S.~Liu, J.~Guo, L.~M. Ni, and L.~Zhang, ``{DN-DETR: Accelerate DETR Training by Introducing Query Denoising},'' in \emph{IEEE/CVF Conf. Comput. Vis. Pattern Recognit. (CVPR)}, New Orleans, LA, USA, Jun. 2022, pp. 13\,619--13\,627.

\bibitem{sparse_detr}
B.~Roh, J.~Shin, W.~Shin, and S.~Kim, ``{Sparse DETR: Efficient End-to-End Object Detection with Learnable Sparsity},'' in \emph{Int. Conf. Learn. Represent. (ICLR)}, Virtual, Apr. 2022, pp. 1--23.

\bibitem{stable_dino}
S.~Liu \emph{et~al.}, ``{Detection transformer with stable matching},'' in \emph{IEEE/CVF Conf. Comput. Vis. Pattern Recognit. (CVPR)}, Van Couver, BC, Canada, Jun. 2023, pp. 6491--6500.

\bibitem{mr_detr}
C.-B. Zhang, Y.~Zhong, and K.~Han, ``{Mr. DETR: Instructive Multi-Route Training for Detection Transformers},'' in \emph{IEEE/CVF Conf. Comput. Vis. Pattern Recognit. (CVPR)}, Nashville, TN, USA, Jun. 2025, p. to appear.

\bibitem{deformable_app0}
J.~Liang \emph{et~al.}, ``{Recurrent Video Restoration Transformer with Guided Deformable Attention},'' in \emph{Adv. Neural Inf. Process. Syst. 35 (NeurIPS)}, New Orleans, LA, USA, 2022, pp. 378--393.

\bibitem{deformable_app1}
W.~Wang \emph{et~al.}, ``{VisionLLM: Large Language Model is also an Open-Ended Decoder for Vision-Centric Tasks},'' in \emph{Adv. Neural Inf. Process. Syst. 36 (NeurIPS)}, New Orleans, LA, USA, 2023, pp. 61\,501--61\,513.

\bibitem{deform_slow0}
P.~Dong \emph{et~al.}, ``{SpeedDETR: Speed-aware Transformers for End-to-end Object Detection},'' in \emph{40th Int. Conf. Mach. Learn. (ICML)}, Honolulu, Hawaii, USA, 2023.

\bibitem{defa}
Y.~Xu \emph{et~al.}, ``{DEFA: Efficient Deformable Attention Acceleration via Pruning-Assisted Grid-Sampling and Multi-Scale Parallel Processing},'' in \emph{61st Annu. Des. Autom. Conf. (DAC)}, San Francisco, CA, USA, Jun. 2024.

\bibitem{ueda}
K.~Sun, M.~Wang, J.~Zhou, and Z.~Wang, ``{{UEDA}}: {{A Universal And Efficient Deformable Attention Accelerator For Various Vision Tasks}},'' in \emph{30th Asia S. Pac. Des. Autom. Conf. (ASP-DAC)}, Tokyo Japan, Jan. 2025, pp. 163--169.

\bibitem{deform_slow1}
T.~Liang and G.~Zeng, ``{FSH-DETR: An Efficient End-to-End Fire Smoke and Human Detection Based on a Deformable DEtection TRansformer (DETR)},'' \emph{Sensors}, vol.~24, no.~13, 2024.

\bibitem{sparse_transformer}
\BIBentryALTinterwordspacing
R.~Child, S.~Gray, A.~Radford, and I.~Sutskever, ``{Generating Long Sequences with Sparse Transformers},'' \emph{arXiv preprint arXiv:1904.10509}, 2019. [Online]. Available: \url{http://arxiv.org/abs/1904.10509}
\BIBentrySTDinterwordspacing

\bibitem{ssr}
J.~Zhuang \emph{et~al.}, ``{SSR: Spatial Sequential Hybrid Architecture for Latency Throughput Tradeoff in Transformer Acceleration},'' in \emph{ACM/SIGDA Int. Symp. Field Program. Gate Arrays (FPGA)}, Monterey, CA, USA, 2024, p. 55–66.

\bibitem{acceltran}
S.~Tuli and N.~K. Jha, ``{AccelTran: A Sparsity-Aware Accelerator for Dynamic Inference With Transformers},'' \emph{{IEEE} Trans. Comput.-Aided Design Integr. Circuits Syst.}, vol.~42, no.~11, pp. 4038--4051, 2023.

\bibitem{hgpipe}
Q.~Guo, J.~Wan, S.~Xu, M.~Li, and Y.~Wang, ``{HG-PIPE: Vision Transformer Acceleration with Hybrid-Grained Pipeline},'' in \emph{IEEE/ACM Int. Conf. Comput. Aided Des. (ICCAD)}, Newark, NJ, USA, Oct. 2024, p. to appear.

\bibitem{random2}
C.~Chen, L.~Li, and M.~M. Sabry~Aly, ``{ViTA: A Highly Efficient Dataflow and Architecture for Vision Transformers},'' in \emph{Des. Autom. Test Eur. (DATE)}, Valencia, Spain, 2024, pp. 1--6.

\bibitem{random3}
W.~Li, Y.~Luo, and S.~Yu, ``{RAWAtten: Reconfigurable Accelerator for Window Attention in Hierarchical Vision Transformers},'' in \emph{Des. Autom. Test Eur. (DATE)}, Antwerp, Belgium, 2023, pp. 1--6.

\bibitem{random0}
E.~Kwon, H.~Song, J.~Park, and S.~Kang, ``{Mobile Accelerator Exploiting Sparsity of Multi-Heads, Lines, and Blocks in Transformers in Computer Vision},'' in \emph{Des. Autom. Test Eur. (DATE)}, Antwerp, Belgium, 2023, pp. 1--6.

\bibitem{random1}
Y.~Zhai, B.~Li, B.~Yan, and J.~Wang, ``{STAR: An Efficient Softmax Engine for Attention Model with RRAM Crossbar},'' in \emph{Des. Autom. Test Eur. (DATE)}, Antwerp, Belgium, 2023, pp. 1--2.

\bibitem{chisel}
J.~Bachrach \emph{et~al.}, ``{Chisel: constructing hardware in a {Scala} embedded language},'' in \emph{49th Annu. Des. Autom. Conf. (DAC)}, San Francisco, CA, USA, Jun. 2012, pp. 1216--1225.

\bibitem{topo}
\BIBentryALTinterwordspacing
``{DC Ultra: Concurrent Timing, Area, Power, and Test Optimization},'' Synopsys, Tech. Rep., 2018. [Online]. Available: \url{https://www.synopsys.com/content/dam/synopsys/implementation%26signoff/datasheets/dc-ultra-ds.pdf}
\BIBentrySTDinterwordspacing

\bibitem{hbm2}
S.~Ghodrati, H.~Sharma, C.~Young, N.~S. Kim, and H.~Esmaeilzadeh, ``{Bit-Parallel Vector Composability for Neural Acceleration},'' in \emph{57th Annu. Des. Autom. Conf. (DAC)}, Virtual, Jul. 2020, pp. 1--6.

\bibitem{coco}
T.-Y. Lin \emph{et~al.}, ``{Microsoft COCO: Common Objects in Context},'' in \emph{13th Eur. Conf. Comput. Vis. (ECCV)}, Zurich, Switzerland, 2014, pp. 740--755.

\bibitem{ppa_scaling}
W.~Huang, K.~Rajamani, M.~R. Stan, and K.~Skadron, ``Scaling with {{Design Constraints}}: {{Predicting}} the {{Future}} of {{Big Chips}},'' \emph{IEEE Micro}, vol.~31, no.~4, pp. 16--29, Jul. 2011.

\end{thebibliography}

\end{document}